\documentclass[prb,twocolumn,showpacs]{revtex4}
\usepackage{color,soul}
\usepackage{amsmath}
\usepackage{dcolumn}
\usepackage{graphicx}
\usepackage{bm}
\usepackage{amssymb}
\usepackage{subfigure}
\begin{document}

\title{A tale of two pairs: on the origin of the pseudogap end point in the high-T$_{c}$ cuprate superconductors}
\author{Jianhua Yang and Tao Li}
\affiliation{Department of Physics, Renmin University of China, Beijing 100872, P.R.China}

\begin{abstract}
There are two seemingly unrelated puzzles about the cuprate superconductors. The first puzzle concerns the strong non-BCS behavior around $x_{c}$, the end point of the superconducting dome on the overdoped side, where the cuprate is believed to be well described by the fermi liquid theory. This is the most evident in the observed $\rho_{s}(0)-T_{c}$ scaling and the large amount of uncondensed optical spectral weight at low energy. The second puzzle concerns the remarkable robustness of the d-wave pairing against the inevitable disorder effect in such a doped system, which is also totally unexpected from the conventional BCS picture. Here we show that these two puzzles are deeply connected to the origin of a third puzzle about the cuprate superconductors, namely, the mysterious quantum critical behavior observed around $x^{*}$, the so called pseudogap end point. Through a systematic variational Monte Carlo(VMC) study of the disordered 2D $t-J$ model from the resonating valence bond(RVB) perspective, we find that the d-wave pairing in this model is remarkably more robust against the disorder effect than that in a conventional d-wave BCS superconductor. We find that such remarkable robustness can be attributed to the spin-charge separation mechanism in the RVB picture, through which the d-wave RVB pairing of the charge neutral spinons becomes essentially immune to the disorder potential except for the secondary effect related to the modulation of the local doping level by the disorder. We propose that there exists a Mott transition at $x^{*}$, where the RVB pairing in the underdoped regime is transmuted into the increasingly more BCS-like pairing for $x>x^{*}$, whose increasing fragility against the disorder effect leads to the non-BCS behavior and the ultimate suppression of superconductivity around $x_{c}$.
\end{abstract}

\maketitle

\section{Introduction}
It is generally believed that the cuprate superconductors will evolve gradually into the conventional BCS superconductors when the doping concentration becomes sufficiently large\cite{Taillefer,Proust,Hussey}. This is supported by the ARPES observation of an increasingly more coherent quasiparticle excitation around the large closed Fermi surface expected from the band theory picture as we increase the doping in the overdoped regime. Results from the quantum oscillation and the Hall response measurement are also consistent with such an understanding. However, recent transport measurements in the heavily overdoped side of the cuprate phase diagram cast serious doubt on such a belief\cite{Bozovic,Mahmood}. In particular, the zero temperature superfluid density $\rho_{s}(0)$ of the cuprate superconductor is found to follow a non-monotonic evolution with the hole concentration, with a prominent peak around the so called pseudogap end point\cite{Uemura,Hussey} $x^{*}\approx0.2$. Quantum critical behaviors of unknown origin are found\cite{Proust} in the low temperature specific heat, dc resistivity and the Hall response around $x^{*}$. In the overdoped side of the phase diagram, $\rho_{s}(0)$ is found to decrease with the increase of hole concentration and vanish with the superconducting critical temperature at the end point of the superconducting dome\cite{Bozovic}($x_{c}\approx0.26$ in La-214 system), in stark contrast to the continuing increase of the Drude weight with doping found in optical measurement\cite{Michon}. This is a totally unexpected behavior for conventional BCS superconductors, in which we expect all Drude weight should be transformed into the superfluid density at zero temperature in the clean limit\cite{Kogan}.

Some people argue that the non-BCS behavior around $x_{c}$ may indicate the potential role of an underlying quantum critical point separating the superconducting phase and an unknown competing phase. For example, a ferromagnetic quantum critical point is proposed in a recent experiment\cite{Kurashima} and it is argued\cite{Mubhoff} that the dominant spin fluctuation would transform from antiferromagnetic to ferromagnetic around $x_{c}$. However, most researchers believe that the non-BCS behavior around $x_{c}$ should be attributed to the inevitable disorder effect related to the dopant out of the $CuO_{2}$ plane\cite{Hone1,Hone,Lizx,Broun,Berg}. In this scenario, the suppression of superconductivity at $x_{c}$ is attributed to the pair breaking effect of the impurity potential in a conventional d-wave superconductor. Indeed, in a recent study of the La-214 system in which the Sr dopant is replaced by Ca to reduce the disorder level of the system(as a result of better match of the Ca dopant ion radius in the system), it is found the end point of the superconducting dome may be extended to substantially higher doping level\cite{Kim}, for example, to doping as high as $x=0.5$.

The effect of disorder in a conventional d-wave BCS superconductor has been studied extensively with both the self-consistent T-matrix theory in the continuum limit at weak disorder level\cite{Markowitz,Alloul,Atkinson,Hone1} and the Bogoliubov-de Gennes mean field theory(BdG) on a lattice\cite{Xiang,Atkinson1,Lizx}. It is found that distinct from the situation in a s-wave superconductor, in which nonmagnetic disorder potential hardly affect the pairing amplitude(the Anderson's theorem)\cite{Anderson}, the d-wave pairing is extremely fragile against the introduction of the nonmagnetic disorder potential. Such a sensitivity can be resorted to the destructive interference of the d-wave pairing amplitude during the impurity scattering process. At strong disorder level, the BdG theory predicts strong spatial inhomogeneity in the pairing amplitude and the emergence of puddles with strong pairing amplitude. Such strong pairing puddles are immersed in a matrix with much reduced pairing amplitude\cite{Lizx}. Exactly such puddling behavior is observed recently in heavily overdoped cuprates around $x_{c}$ through scanning microscopic spectroscopy\cite{Tromp}. The weak links between such sparse strong pairing puddles are argued to be responsible for the strong phase fluctuation effect around $x_{c}$. This may offer a consistent interpretation for the $T_{c}-\rho_{s}(0)$ scaling and the large amount of uncondensed low energy spectral weight in the optical spectrum observed around $x_{c}$.

On the other hand, the superconducting state in the underdoped cuprates seems to be remarkably robust against the out-of-plane disorder caused by the dopants, even though it possess exactly the same d-wave symmetry\cite{Alloul,Kim}. This is strikingly distinct from our understanding based on the standard BCS scenario. It was proposed early on that the strong correlation effect inherent in the cuprates may help to enhance the robustness of the d-wave superconductivity against the out-of-plane disorder\cite{Grag,Chakra,Ghosal}. Based on a BdG treatment of the $t-J$ Hamiltonian supplemented by a Gutzwiller approximation of the no double occupancy constraint in the 2D $t-J$ model(BdG+GA), the authors of Ref.[\onlinecite{Grag}] argued that the spatial variation in the Gutzwiller factor of the hopping term acts to compensate the effect of the disorder potential. Here the Gutzwiller factor is introduced to approximate the effect of no double occupancy constraint in the $t-J$ model. However, an exact treatment of such local constraint in the presence of the disorder potential is absent.

Here we perform a systematic variational Monte Carlo study of the disorder effect in the 2D $t-J$ model based on the resonating valence bond(RVB) theory\cite{RVB,PALee}, in which the no double occupancy constraint is treated exactly. We model the disorder effect caused by out-of-plane dopants with a random on-site impurity potential. Similar to what is found in previous BdG+GA treatment, we find that the d-wave pairing in the 2D $t-J$ model is remarkably more robust against the disorder effect than that in a conventional d-wave BCS superconductor. For example, we find that the reduction in the off-diagonal-long-range-order(ODLRO) in the presence of the impurity potential never exceed 20 percent of its clean limit value at any doping, even if the impurity potential is more than 8 times stronger than the Heisenberg exchange coupling $J$ in the $t-J$ model. We find that such remarkable robustness of the d-wave pairing can be attributed to the spin-charge separation mechanism in the RVB picture, through which the d-wave RVB pairing of the charge neutral spinons becomes essentially immune to the impurity potential. Indeed, we find that it is the holon degree of freedom that is most significantly affected by the impurity potential and that the impurity effect on the spinon part is much reduced. For example, the spatial variation in the spinon chemical potential is found to be an order of magnitude smaller than that in the bare impurity potential. This forms a strong contrast with the situation in a disordered d-wave BCS superconductor, in which the impurity potential acts directly on the electron participating in the d-wave pairing.

Based on these results, we propose the following scenario for the non-monotonic doping dependence of $\rho_{s}(0)$. For $x<x^{*}$, where the cuprate superconductor is well described by a doped Mott insulator, the d-wave pairing should be better understood as the RVB pairing between charge neutral spinons which is essentially immune to the impurity potential. $\rho_{s}(0)$ in this regime should be dominated by the density of mobile charge carriers and should thus increase monotonically with $x$, as what we expect for a doped Mott insulator. For $x>x^{*}$, a description in terms of the conventional fermi liquid metal becomes increasingly more relevant with the increase of the doping level. The RVB pairing in the underdoped regime will be gradually transmuted into the conventional BCS pairing that is fragile in the presence of impurity scattering. This explains the increasing fragility of the d-wave pairing against the disorder effect and the ultimate suppression of superconductivity at $x_{c}$ in the overdoped regime. The superfluid density is expected to decrease with increasing doping as a result of such transmutation in the nature of the electron in the system, namely the transmutation from local moments to itinerant quasiparticles.

 A sharp transition at $x^{*}$ from a doped Mott insulator to a less correlated fermi liquid metal is implicitly assumed in the above scenario. While Mott transition at a general incommensurate filling is still not a well accepted notion, several measurements are consistent with the abrupt enhancement of electron itineracy at $x^{*}$\cite{Tallon,Davis,Minola,Chen}. In addition, $x^{*}$ is also found to be position where field induced spin glass behavior starts to emerge\cite{Julien}. Theoretically, such a transition has been claimed in a DMFT study of the Hubbard model\cite{Sordi}. In the real cuprate superconductors, both the feedback effect of the enhanced electron itineracy on the screening of the Coulomb repulsion and the reduction in the charge transfer gap with hole doping may play important role in driving such a transition. As $x^{*}$ is also the doping where the strange metal behavior becomes the most evident and that pseudogap phenomena starts to emerge, we think that these two major mysteries of the cuprate physics should all be attributed to such a finite doping Mott transition. Such a transition is surely beyond the Landau paradigm of conventional quantum phase transition which involves spontaneous symmetry breaking order. In particular, we think the Mottness of electron and the RVB pairing between the charge neutral spinons in such a doped Mott insulator is at the heart of the origin of the enigmatic pseudogap phenomena.

The paper is organized as follows. In the next section, we describe the disordered 2D $t-J$ model and its variational ground state. In the third section we discuss the optimization of the variational parameters in the RVB state constructed in section II. The fourth section is devoted to the presentation of the results we got from the VMC calculation. We draw conclusion from these numerical results in the last section and discuss their implication on the physics of the cuprate superconductors.

\section{The disordered 2D $t-J$ model and its variational ground state}
The model we study in this work is described by the following Hamiltonian
\begin{eqnarray}
H&=&-t\sum_{\langle i,j \rangle,\alpha}(\hat{c}^{\dagger}_{i,\alpha}\hat{c}_{j,\alpha}+h.c.)+\sum_{i,\alpha}\mu_{i} \hat{c}^{\dagger}_{i,\alpha}\hat{c}_{i,\alpha}\nonumber\\
&-&t'\sum_{\langle\langle i,j \rangle\rangle,\alpha}(\hat{c}^{\dagger}_{i,\alpha}\hat{c}_{j,\alpha}+h.c.)\nonumber\\
&+&J\sum_{\langle i,j \rangle}(\mathbf{S}_{i}\cdot\mathbf{S}_{j}-\frac{1}{4}n_{i}n_{j})
\end{eqnarray} 
here $t$ and $t'$ denote the hopping integral of electron between nearest-neighboring(NN) and next-nearest-neighboring(NNN) sites. $\hat{c}_{i,\sigma}$ denotes the electron annihilation operator on site $i$ and with spin $\alpha$. It should satisfy the following constraint of no double occupancy
\begin{equation}
\sum_{\alpha}\hat{c}^{\dagger}_{i,\alpha}\hat{c}_{i,\alpha}\leq 1
\end{equation}
$J$ denotes the Heisenberg exchange coupling between NN sites. The last term in the first line represents the disorder effect of a random onsite potential $\mu_{i}$. We choose a box distribution for $\mu_{i}$ in this study, namely, $\mu_{i}$ is distributed uniformly in a box region $[-\frac{V}{2},\frac{V}{2}]$.  In our study, we set $t'/t=-0.3$ and $J/t=0.3$. We will use $t$ as the unit of energy. The disorder strength $V/t$ is varied in the range of $V/t\in[0,2.5]$. Such a disorder strength is strong enough to kill the d-wave superconducting pairing totally if we ignore the electron correlation effect, namely, if we ignore the no double occupancy constraint on the electron operator.

Here we study the ground state property of the disordered 2D $t-J$ model with the variational Monte Carlo(VMC) method within the fermionic RVB framework. To motivate the form of the variational ground state, we represent the constrained electron operator $\hat{c}_{i,\sigma}$ in terms of the charge neutral spinon operator $f_{i,\sigma}$ and the spinless holon operator $b_{i}$ as follows
 \begin{equation}
 \hat{c}_{i,\sigma}=f_{i,\sigma}b^{\dagger}_{i}
 \end{equation}
Here $f_{i,\sigma}$ and $b_{i}$ are fermionic and bosonic operators. This is an exact representation of the constrained electron operator if the the spinon and the holon operator satisfy the constraint 
\begin{equation}
\sum_{\alpha}f^{\dagger}_{i,\alpha}f_{i,\alpha}+b^{\dagger}_{i}b_{i}=1
\end{equation}
The RVB variational ground state we adopted to describe the ground state of the disordered 2D $t-J$ model takes the form of
\begin{equation}
|\mathrm{RVB}\rangle=\mathrm{P}_{\mathrm{G}}| f-\mathrm{BCS}\rangle\otimes |b-\mathrm{Condens}\rangle
\end{equation}
in which $\mathrm{P}_{\mathrm{G}}$ denotes the Gutzwiller projection enforcing the no double occupancy constraint Eq.4. $| f-\mathrm{BCS}\rangle$ denotes the mean field ground state of the fermionic spinon in the slave boson mean field theory of the disordered 2D $t-J$ model. $|b-\mathrm{Condens}\rangle$ denotes the wave function of holon condensate on such a disordered background. We emphasis that the no double occupancy constraint in the $t-J$ model is crucial to arrive at our conclusion and is treated exactly in the variational approach adopted in this work.

To derive the detailed form of both $| f-\mathrm{BCS}\rangle$ and $|b-\mathrm{Condens}\rangle$, we rewrite the Hamiltonian of the disordered 2D $t-J$ model in terms of the spinon and the holon operator as follows, 
\begin{eqnarray}
H=&-&t\sum_{\langle i,j\rangle,\alpha}(f^{\dagger}_{i,\alpha}f_{j,\alpha}b^{\dagger}_{i}b_{j}+h.c.)+\sum_{i,\alpha}\mu_{i}f^{\dagger}_{i,\alpha}f_{i,\alpha}\nonumber\\
&-&t'\sum_{\langle\langle i,j\rangle\rangle,\alpha}( f^{\dagger}_{i,\alpha}f_{j,\alpha}b^{\dagger}_{i}b_{j}+h.c.)\nonumber\\
&+&\frac{J}{2}\sum_{\langle i,j \rangle,\alpha,\beta}[f^{\dagger}_{i,\alpha}f_{i,\beta}f^{\dagger}_{j,\beta}f_{j,\alpha}-f^{\dagger}_{i,\alpha}f_{i,\alpha}f^{\dagger}_{j,\beta}f_{j,\beta}]\nonumber\\
\end{eqnarray} 
in which $\alpha,\beta=\uparrow,\downarrow$. Here we have represented the spin operator as 
\begin{equation}
\mathbf{S}_{i}=\frac{1}{2}\sum_{\alpha,\beta}f^{\dagger}_{i,\alpha}\bm{\sigma}_{\alpha,\beta}f_{i,\beta}
\end{equation}
in which $\bm{\sigma}$ is the usual Pauli matrix for electron spin. The electron number operator is represented as
\begin{equation}
n_{i}=\sum_{\alpha}f^{\dagger}_{i,\alpha}f_{i,\alpha}
\end{equation}
We then perform the mean field decoupling of the Hamiltonian by introducing the following RVB order parameters for the spinon  
\begin{eqnarray}
\chi_{i,j}&=&\langle f^{\dagger}_{i,\uparrow}f_{j,\uparrow}+f^{\dagger}_{i,\downarrow}f_{j,\downarrow}\rangle\nonumber\\
\Delta_{i,j}&=&\langle f^{\dagger}_{i,\uparrow}f^{\dagger}_{j,\downarrow}+f^{\dagger}_{j,\uparrow}f^{\dagger}_{i,\downarrow}\rangle
\end{eqnarray}
and the boson condensate amplitude for the holon
\begin{equation}
\bar{b}_{i}=\langle b_{i} \rangle=\langle b^{\dagger}_{i} \rangle
\end{equation}
In this study we will assume $\chi_{i,j}$, $\Delta_{i,j}$ and $\bar{b}_{i}$ to be real numbers. After the mean field decoupling, the spinon Hamiltonian takes the form of 
\begin{eqnarray}
H^{f}_{MF}=&-&\sum_{\langle i,j\rangle,\alpha}t^{v}_{i,j}(f^{\dagger}_{i,\alpha}f_{j,\alpha}+h.c.)+\sum_{i,\alpha}\mu^{v}_{i}f^{\dagger}_{i,\alpha}f_{i,\alpha}\nonumber\\
&-&\sum_{\langle\langle i,j\rangle\rangle,\alpha}t'^{v}_{i,j}( f^{\dagger}_{i,\alpha}f_{j,\alpha}+h.c.)\nonumber\\
&+&\sum_{\langle i,j \rangle}\Delta^{v}_{i,j}(f^{\dagger}_{i,\uparrow}f^{\dagger}_{j,\downarrow}+f^{\dagger}_{j,\uparrow}f^{\dagger}_{i,\downarrow}+h.c.)
\end{eqnarray}
$| f-\mathrm{BCS}\rangle$ used to construct the variational ground state $|\mathrm{RVB}\rangle$ can be obtained by diagonalizing numerically such a Bogliubov-de Gennes Hamiltonian on a finite lattice. At the same time, the holon mean field ground state takes the form of 
\begin{equation}
|b-\mathrm{Condens}\rangle=\left(\sum_{i}\bar{b}_{i}b^{\dagger}_{i}\right)^{N_{b}}|0\rangle_{b}
\end{equation}
Here $N_{b}$ is the number of doped holes, $|0\rangle_{b}$ denotes the vacuum of the holon Hilbert space.

We emphasize again that the mean field state $|f-\mathrm{BCS}\rangle$ and $|b-\mathrm{Condens}\rangle$ are just intermediate steps used to construct the physical variational ground state $|\mathrm{RVB}\rangle$, in which the no double occupancy constraint is treated exactly. The variational parameter involved in $|\mathrm{RVB}\rangle$ includes $t^{v}_{i,j}$, $t'^{v}_{i,j}$, $\Delta^{v}_{i,j}$, $\mu^{v}_{i}$ appearing in Eq.11 and $\bar{b}_{i}$ appearing in Eq.12. Note that in the presence of the disorder potential, these variational parameters must be assumed to be spatial inhomogeneous and be optimized independently.

\section{Variational optimization of the RVB wave function}
To optimized the variational parameters involved in $|\mathrm{RVB}\rangle$, we expand it in a orthornomal basis $|R\rangle$ and denote the corresponding wave function amplitude as $\Psi(R)$. The variational ground state energy is then given by
\begin{equation}
E=\langle H \rangle_{\Psi}=\frac{\langle\Psi| H |\Psi\rangle}{\langle \Psi |\Psi \rangle}=\frac{\sum_{R}|\Psi(R)|^2 E_{loc}(R)}{\sum_{R}|\Psi(R)|^2}
\end{equation}
in which the local energy $E_{loc}(R)$ is defined as
\begin{equation}
E_{loc}(R)=\sum_{R'}\langle R |H| R' \rangle \frac{\Psi(R')}{\Psi(R)}
\end{equation}
The gradient of the variational energy with respect to the variational parameters is given by
\begin{equation}
\nabla E= \langle \nabla \ln \Psi(R) E_{loc}(R) \rangle_{\Psi}-E\langle \nabla \ln \Psi(R) \rangle_{\Psi}
\end{equation}
Here we denote the variational parameters collectively as $\bm{\alpha}$ and abbreviate $\nabla_{\bm{\alpha}}$ as $\nabla$. Both $E$ and $\nabla E$ can be computed by standard Monte Carlo sampling over the distribution generated by $|\Psi(R)|^2$. 

We now derive an expression for $\Psi(R)$. For this purpose we rewrite the spinon Hamiltonian in the following form
\begin{equation}
H^{f}_{MF}=\psi^{\dagger}\mathbf{M}\psi
\end{equation}
in which 
\begin{equation}
\psi^{\dagger}=(f^{\dagger}_{1,\uparrow},...,f^{\dagger}_{N,\uparrow},f_{1,\downarrow},....,f_{N,\downarrow})
\end{equation}
Here $N$ denotes the total number of lattice site and $\mathbf{M}$ is a $2N\times 2N$ Hermitian matrix. For computational convenience, we make the following unitary transformation on the spinon operator
\begin{eqnarray}
f_{i,\uparrow}& \rightarrow &\tilde{f}_{i,\uparrow}\nonumber\\
f_{i,\downarrow}&\rightarrow&\tilde{f}^{\dagger}_{i,\downarrow}
\end{eqnarray}
The mean field ground state of $H^{f}_{MF}$ is then constructed by filling up all eigenstates with negative eigenvalue for the $\tilde{f}$-fermion and takes the form of
\begin{equation}
|f-\mathrm{BCS}\rangle=\prod_{n=1}^{N}\gamma^{\dagger}_{n}\ |0\rangle_{\tilde{f}}
\end{equation}
in which 
\begin{equation}
|0\rangle_{\tilde{f}}=\prod_{i=1}^{N}f^{\dagger}_{i,\downarrow}|0\rangle_{f}
\end{equation}
is the vacuum state of the $\tilde{f}$-fermion and $|0\rangle_{f}$ is the vacuum state of the original $f$-fermion. $\gamma^{\dagger}_{n}$ denotes the creation operator of the $n$-th Bogliubov quasiparticle with a negative eigenvalue and is given by
\begin{equation}
\gamma^{\dagger}_{n}=\sum^{N}_{i=1}[\phi_{n}(i)\tilde{f}^{\dagger}_{i,\uparrow}+\phi_{n}(i+N)\tilde{f}^{\dagger}_{i,\downarrow}]
\end{equation}
in which $\phi_{n}(i)$ denotes the $i$-th component of the $n$-th eigenvector of the matrix $\mathbf{M}$.

A general basis vector satisfying the no double occupancy constraint between the spinon and the holon can be written as
\begin{equation}
|R\rangle=\prod^{N_{e}}_{k=1}\tilde{f}^{\dagger}_{i_{k},\uparrow}\prod^{N-N_{e}}_{k'=1}\tilde{f}^{\dagger}_{j_{k'},\downarrow}|0\rangle_{\tilde{f}}\otimes \prod^{N_{b}}_{k''=1}b^{\dagger}_{l_{k''}}|0\rangle_{b}
\end{equation}
in which $i_{k}$ denotes the position of the $k$-th up-spin electrons, $j_{k'}$ denotes the position of the $k'$-th hole of the down-spin electrons, $l_{k''}$ denotes the position of the $k''$-th unoccupied sites. Note that we have equal number of up-spin and down-spin electron($N_{e}$) in the ground state. The no-double-occupancy constraint requires that 
\begin{equation}
2N_{e}+N_{b}=N
\end{equation} 
At the same time, each of the $N-N_{e}$ sites occupied by the $\tilde{f}_{\downarrow}$ fermion should either be simultaneously occupied by a $\tilde{f}_{\uparrow}$ fermion or left empty.
The wave function amplitude $\Psi(R)$ of $|\mathrm{RVB}\rangle$ in this basis is given by
\begin{equation}
\Psi(R)=\mathrm{Det}[\bm{\Phi}]\times \prod_{k=1}^{N_{b}} \bar{b}_{l_{k}}
\end{equation}
in which $\bm{\Phi}$ is a $N\times N$ matrix of the form
\begin{eqnarray}
\bm{\Phi}=\left(\begin{array}{cccc}\phi_{1}(i_{1}) & . & . & \phi_{N}(i_{1}) \\. & . & . & . \\ \phi_{1}(i_{N_{e}}) & . & . & \phi_{N}(i_{N_{e}}) \\   \phi_{1}(j_{1}) & . & . & \phi_{N}(j_{1}) \\. & . & . & .\\
 \phi_{1}(j_{N-N_{e}}) & . & . & \phi_{N}(j_{N-N_{e}})\nonumber
 \end{array}\right)
\end{eqnarray}

In the calculation of the energy gradient, the key quantity to be computed is $\nabla \ln \Psi(R)$. For the variational parameters appearing in $H^{f}_{MF}$, we have
\begin{equation}
\nabla\ln \Psi(R)=\mathbf{Tr} [\nabla \bm{\Phi} \bm{\Phi}^{-1}]
\end{equation}
The matrix elements of $\nabla \bm{\Phi}$ can be calculated from the first order perturbation theory as follows
\begin{equation}
\nabla \phi_{n}=\sum_{E_{m}>0}\frac{\langle\phi_{m}|\nabla H^{f}_{MF}|\phi_{n}\rangle}{E_{n}-E_{m}}\phi_{m}
\end{equation}  
Here $|\phi_{n} \rangle$ and $E_{n}$ denote the $n$-th eigenvector and eigenvalue of the mean field Hamiltonian $H^{f}_{MF}$. The gradient of $\Psi(R)$ in the holon condensation parameter $\bar{b}_{i}$ is given by
\begin{equation}
\frac{\partial}{\partial \bar{b}_{i}}\ln \Psi(R)=\frac{1}{\bar{b}_{i}}\sum^{N_{b}}_{k=1}\delta_{l_{k},i}
\end{equation}

On a finite cluster of square lattice with $N$ sites, there are in total $8N$ variational parameters to be optimized. These include $2N$ NN hopping parameter $t^{v}_{i,j}$, $2N$ NNN hopping parameter $t'^{v}_{i,j}$, $2N$ NN pairing parameter $\Delta^{v}_{i,j}$, $N$ on-site parameter $\mu^{v}_{i}$ and $N$ holon condensate amplitude $\bar{b}_{i}$. The efficient optimization of such a large number of variational parameters constitutes a big challenge for variational Monte Carlo method. In a recent work, we have proposed several improved algorithms to achieve such a goal\cite{Li}. These include the traditional steepest descent(SD) or the stochastic reconfiguration(SR) method accelerated by a self-learning trick and a finite-depth realization of the BFGS or the conjugate gradient(CG) method. Here we will adopt the steepest descent method accelerated by the self-learning trick. The SD algorithm is the simplest optimization algorithm. It corresponds to setting the Hessian matrix of the problem proportional to the identity matrix. In the SD algorithm, the variational parameters are updated as follows
\begin{equation}
\bm{\alpha}\rightarrow \bm{\alpha}-\delta \nabla E
\end{equation} 
 in which $\delta$ is the step length. Instead of using a fixed step length, here we adjust $\delta$ adaptively according to the following rule
 \begin{equation}
 \delta\rightarrow \delta\times(1+ \eta \frac{\nabla E\cdot \nabla E'}{|\nabla E||\nabla E'|})
 \end{equation}  
 in which $\eta\in [0,1]$ is an acceleration factor, $\nabla E$ and $\nabla E'$ is the gradient of the energy in the current and the previous step. According to such a rule, the step length will be enhanced(reduced) if the successive energy gradients tend to point to the same(opposite) direction. In our calculation we will set $\eta=1$.

 As a result of the heavy computational cost in the variational optimization of such large number of variational parameters, we will focus on a fixed realization of the impurity potential $\mu_{i}$. More specifically, we set
\begin{equation}
\mu_{i}=V(r_{i}-0.5)
\end{equation}
in which $r_{i}$ is random number uniformly distributed in the range of $[0,1]$. We will fix $r_{i}$ and tune the disorder strength by varying $V$. Since most of the quantities that concern us are subjected to self-averaging, the absence of the disorder average does not cause any essential problem. For example, the off-diagonal-long-range-order calculated from such a fixed disorder realization setup exhibits smooth evolution with both $V/t$ and the doping concentration(see Fig.8 below).

\section{Numerical results}
We have performed variational optimization of the 2D $t-J$ model using $|\mathrm{RVB}\rangle$ as the variational ground state. The calculation is done on a $12\times12$ lattice with periodic-antiperiodic boundary condition. Since the variational ground state is invariant under a global rescaling of the variational parameters appearing in the spinon Hamiltonian $H^{f}_{MF}$, we will measure the variational parameters in $H^{f}_{MF}$ in unit of $t^{v}_{1,1+x}$, namely, the first NN hopping parameter in the $x$-direction. The calculation is done for 12 hole concentrations corresponding to hole number $N_{b}=12,16,20,24,28,32,34,36,40,44,48,52$, which corresponds to the doping range $x\in[0.083,0.361]$. This covers the most part of the superconducting dome in the cuprate phase diagram. The optimized paring amplitude in the absence of the impurity potential is shown in Fig.1 for reference. As can be seen from the figure, the pairing amplitude is already rather small when $N_{b}=52$, which corresponds to a hole concentration of $x=0.361$.

\begin{figure}
\includegraphics[width=8.5cm]{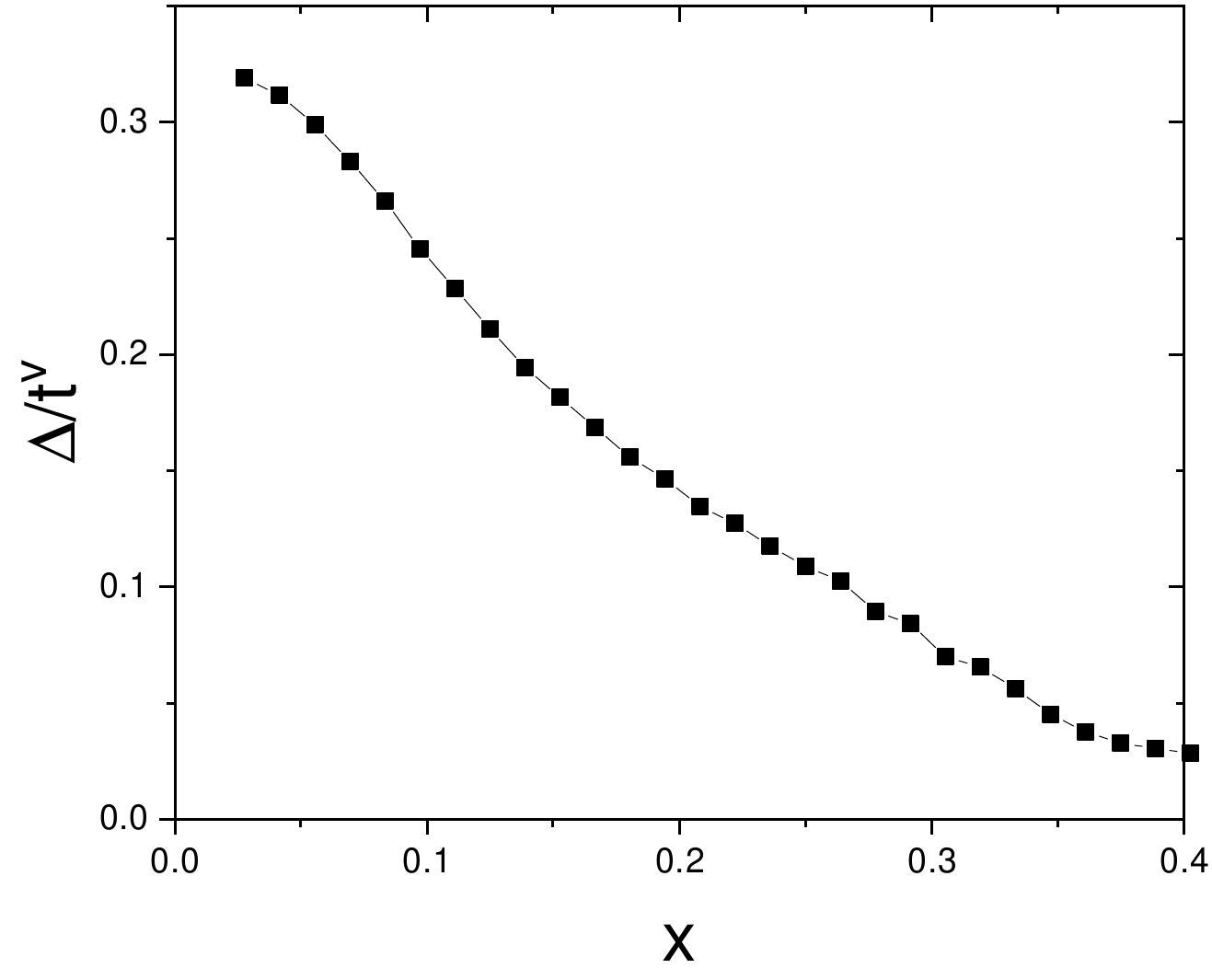}
\caption{The evolution of the optimized pairing amplitude with the doping concentration $x$ in the absence of the impurity potential. The calculation is done on a $12\times12$ lattice with periodic-antiperiodic boundary condition. Here we measure the pairing amplitude in unit of the NN hopping parameter $t^{v}$.}
\end{figure}
\begin{figure}
\includegraphics[width=9.5cm]{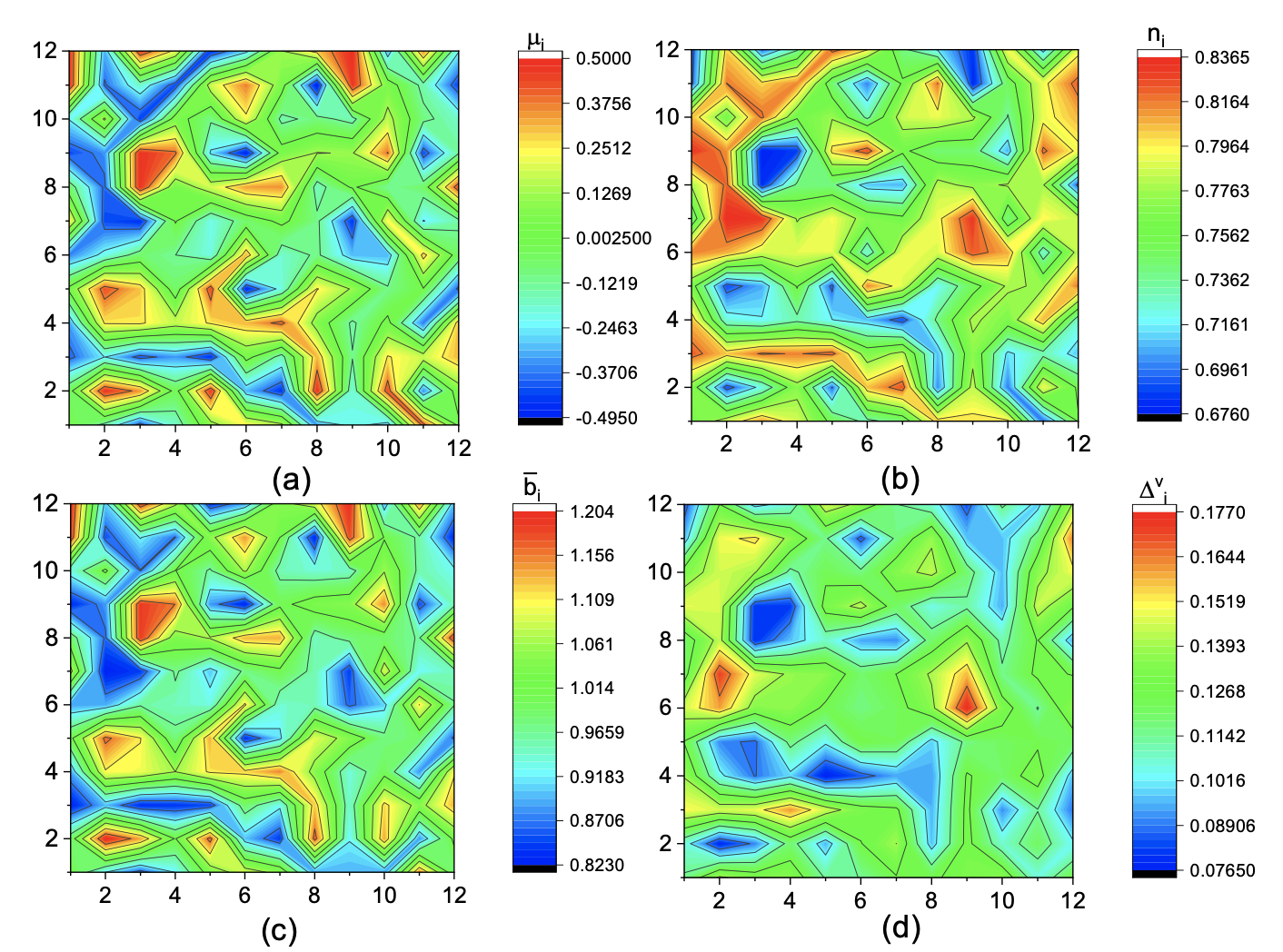}
\caption{The spatial distribution of the optimized holon condensate amplitude $\bar{b}_{i}$(Fig.2c) and the site average of the pairing amplitude $\Delta^{v}_{i}$(Fig.2d) at $x=0.236$ and $V/t=1$ for a particular realization of the disorder potential $\mu_{i}$(Fig.2a). The the optimized local electron density $n_{i}$ is shown in Fig.2b for reference. As is naturally expected, there is strong positive(negative) correlation between the optimized holon condensate amplitude $\bar{b}_{i}$(local electron density $n_{i}$) and the disorder potential $\mu_{i}$. The strong negative correlation between $\Delta^{v}_{i}$ and $\mu_{i}$(or $\bar{b}_{i}$) is also evident from Fig.2d, implying that the pairing amplitude is mainly modulated by the local hole concentration. The calculation is done on a $12\times12$ lattice with periodic-antiperiodic boundary condition.}
\end{figure}

Now we consider the disordered case. We find that the disorder potential has its most significant effect on the Boson condensate amplitude $\bar{b}_{i}$ and the pairing amplitude $\Delta^{v}_{i,j}$. On the other hand, the hopping parameter $t^{v}_{i,j}$, $t'^{v}_{i,j}$ and the local chemical potential parameter $\mu^{v}_{i}$ are found to be only weakly affected. Shown in Fig.2  are the spatial distribution of the optimized holon condensate amplitude and the pairing amplitude at $x=0.236$ and $V/t=1.0$ for a particular realization of the disorder potential $\mu_{i}$. To better visualize the spatial distribution of the pairing amplitude, we have defined the following site average for the bond variable
$\Delta^{v}_{i,j}$
\begin{equation}
\Delta^{v}_{i}=\frac{1}{4}[\Delta^{v}_{i,i+x}+\Delta^{v}_{i,i-x}-\Delta^{v}_{i,i+y}-\Delta^{v}_{i,i-y}]
\end{equation}
We note that the optimized pairing amplitude preserves the d-wave phase structure both locally and globally, namely, the pairing amplitude in the $x$ and the $y$-direction are always opposite in their signs. This is very different from the result of plain BdG calculation, in which the destructive interference of the pairing amplitude may lead to frustration in its phase\cite{Lizx,Berg}.

From Fig.2 it is clear that there is strong positive correlation between the optimized holon condensate amplitude $\bar{b}_{i}$ and the disorder potential $\mu_{i}$. This is a naturally expected result since with the increase of $\mu_{i}$ the local electron density would be depleted. This is indeed found in the optimized local electron density $n_{i}$ shown in Fig2b, which exhibits strong negative correlation with $\mu_{i}$. The optimized pairing amplitude $\Delta^{v}_{i}$ is found to exhibit strong negative correlation with $\mu_{i}$ or $\bar{b}_{i}$, implying that it is mainly modulated by the local hole concentration. It is important to note that while the pairing amplitude $\Delta^{v}_{i}$ is spatially inhomogeneous at the lattice scale, there is no puddling behavior in its distribution even at the strongest disorder potential with $V/t=2.5$, which is more than 8 times stronger than the Heisenberg exchange coupling $J$. This is very different from the result predicted by the plain BdG calculation.

\begin{figure}
\includegraphics[width=8cm]{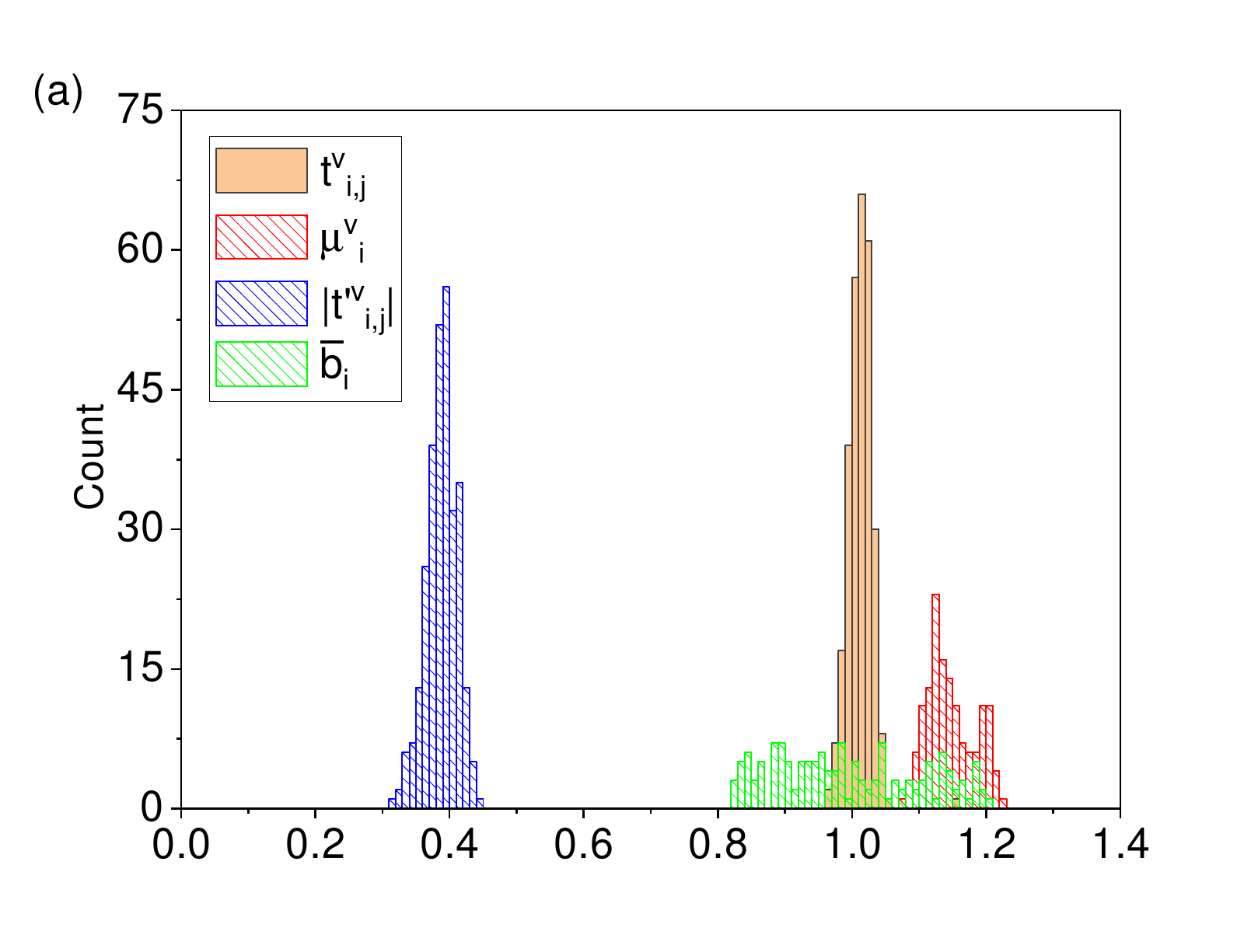}
\includegraphics[width=8cm]{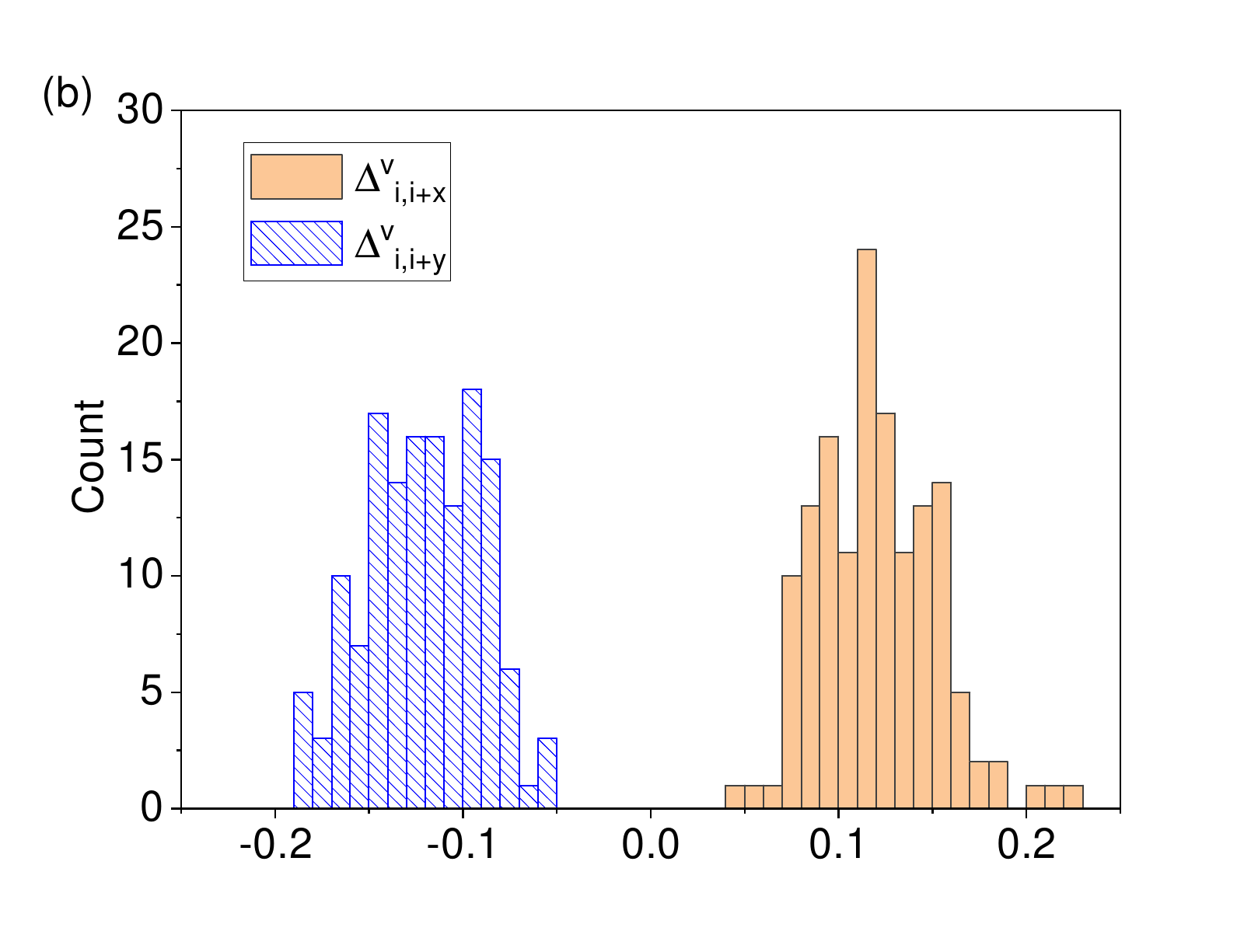}
\caption{The histogram of the distribution in the value of the optimized variational parameters at $x=0.236$ with $V/t=1$. The calculation is done on a $12\times12$ lattice with periodic-antiperiodic boundary condition. Here  $t^{v}_{i}$, $t'^{v}_{i,j}$, $\mu^{v}_{i}$ and $\Delta^{v}_{i,j}$ are all measured in unit of the first NN hopping parameter in the $x$-direction, namely  $t^{v}_{1,1+x}$.}
\end{figure}

To be more quantitative, we plot in Fig.3 the histogram of the distribution in the value of the optimized variational parameters. The standard variation in the fluctuation of the optimized parameters are respectively 
\begin{eqnarray}
\delta \mu^{v}_{i}&\approx&0.03\ \langle \mu^{v}_{i}\rangle\nonumber\\
\delta t^{v}_{i,j}&\approx&0.016\ \langle t^{v}_{i,j}\rangle\nonumber\\
\delta |t'^{v}_{i,j}|&\approx&0.058\ \langle |t'^{v}_{i,j}|\rangle\nonumber\\
\delta |\Delta^{v}_{i,j}|&\approx&0.26\ \langle |\Delta^{v}_{i,j}|\rangle\nonumber\\
\delta \bar{b}_{i}&\approx&0.11\ \langle \bar{b}_{i}\rangle
\end{eqnarray}
Here $\langle\  \rangle$ denotes spatial average. To understand how the disorder potential play its role, we plot in Fig.4 the correlation between the optimized variational  parameters and the bare local chemical potential $\mu_{i}$. Similar to $\Delta^{v}_{i}$, we have defined the site average for the bond variable $t^{v}_{i,j}$ and $t'^{v}_{i,j}$ as follows
\begin{eqnarray}
t^{v}_{i}&=&\frac{1}{4}[t^{v}_{i,i+x}+t^{v}_{i,i-x}+t^{v}_{i,i+y}+t^{v}_{i,i-y}]\nonumber\\
t'^{v}_{i}&=&\frac{1}{4}[t'^{v}_{i,i+x+y}+t'^{v}_{i,i-x-y}+t'^{v}_{i,i+x-y}+t'^{v}_{i,i-x+y}]\nonumber\\
\end{eqnarray}
 Clearly, it is the holon condensate amplitude $\bar{b}_{i}$ that has the strongest correlation with the local chemical potential $\mu_{i}$. On the other hand, the variation of the spinon chemical potential $\mu^{v}_{i}$(measured in unit of $t^{v}_{1,1+x}$, the first NN spinon hopping parameter in the $x$-direction) is an order of magnitude weaker than that in the bare chemical potential $\mu_{i}$(measured in unit of the NN hopping integral of the 2D $t-J$ model). Thus, the 2D $t-J$ model responds to the disorder potential mainly though the deformation of the holon condensate rather than the structure of the spinon ground state. This is also evident in the hopping parameter $t^{v}$ and $t'^{v}$, which exhibit negligible correlation with the bare local chemical potential. The most natural way to understand such a peculiar behavior is through the notion of spin-charge separation in a doped Mott insulator. More specifically, the disorder potential is mainly experienced by the spinless holon which carries the charge of an electron, while the charge neutral spinon responsible for the d-wave RVB pairing is almost immune to the impurity potential. This is very different from the situation in a BCS superconductor, in which it is the electron which participates in the Cooper pairing that experience the disorder potential.

The correlation between the pairing amplitude $\Delta^{v}$ and the local chemical potential $\mu_{i}$ is more subtle. As can be seen form Fig.4b, $\Delta^{v}$ exhibits a clear  anti-correlation with $\mu_{i}$. Such a behavior can be understood as the result of the modulation of the local doping concentration by the disorder potential. More specifically, the magnitude of the pairing amplitude will be suppressed in regions with a higher hole density and be enhanced in regions with a lower hole density. Such a scenario is supported by the correlation between the local hole density and the magnitude of the pairing amplitude shown in Fig.5, which follows nicely the doping dependence of the pairing amplitude in the clean system. This also explain why the d-wave phase structure is well preserved in the disordered system.

\begin{figure}
\includegraphics[width=8cm]{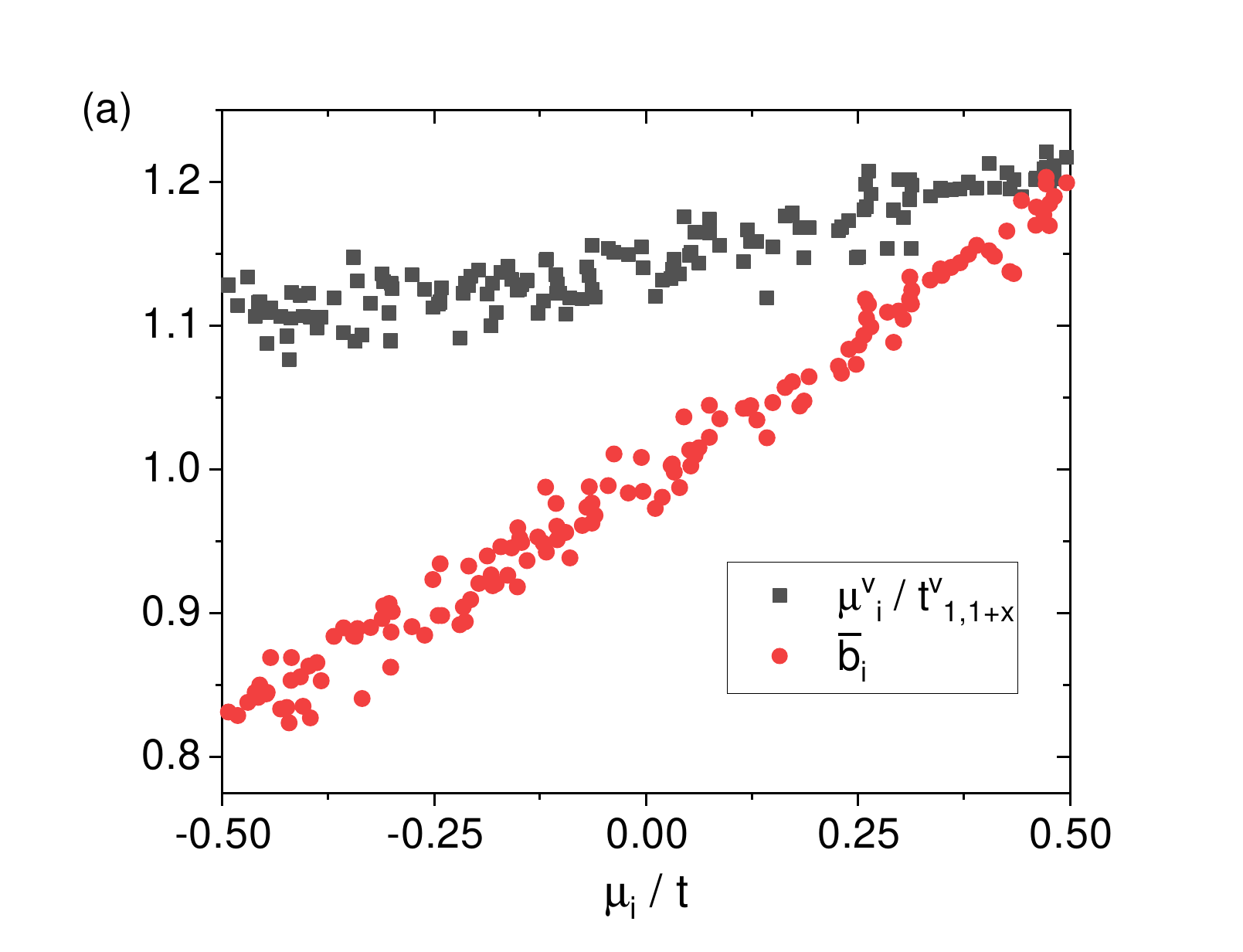}
\includegraphics[width=8cm]{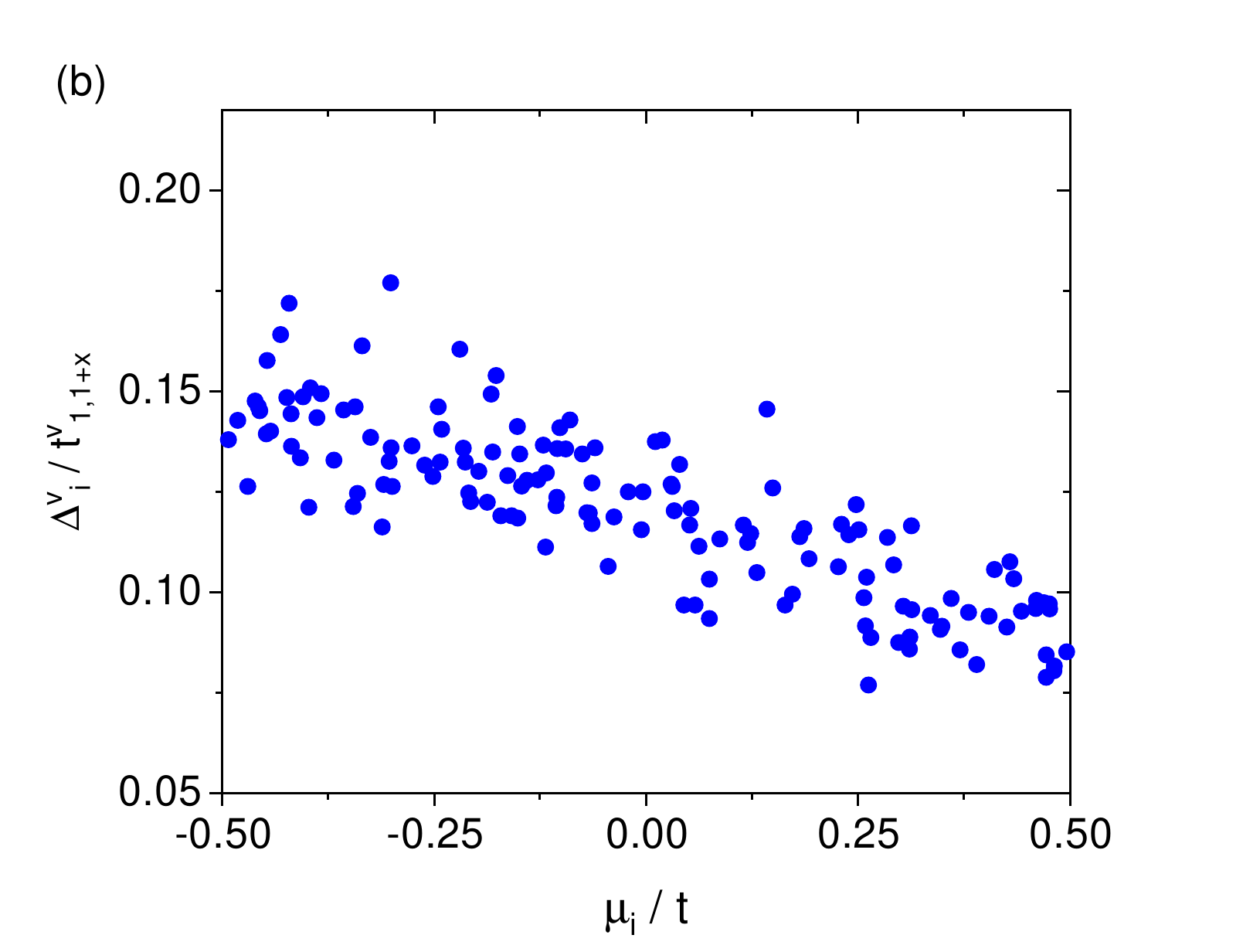}
\includegraphics[width=8cm]{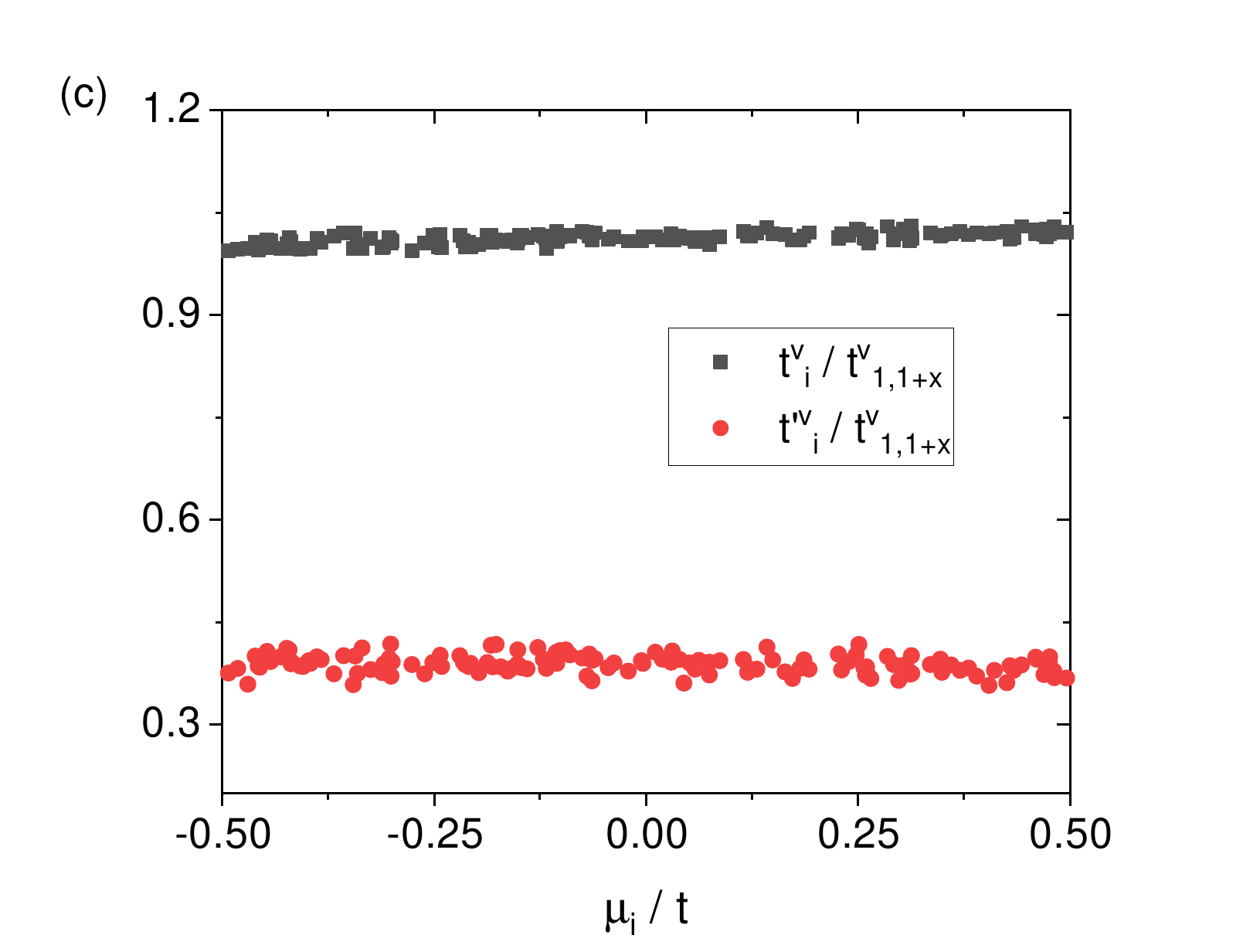}
\caption{The correlation between the optimized variational parameters and the bare local chemical potential $\mu_{i}$ at $x=0.236$. Here we set $V/t=1$. The calculation is done on a $12\times12$ lattice with periodic-antiperiodic boundary condition. }
\end{figure}  

\begin{figure}
\includegraphics[width=8cm]{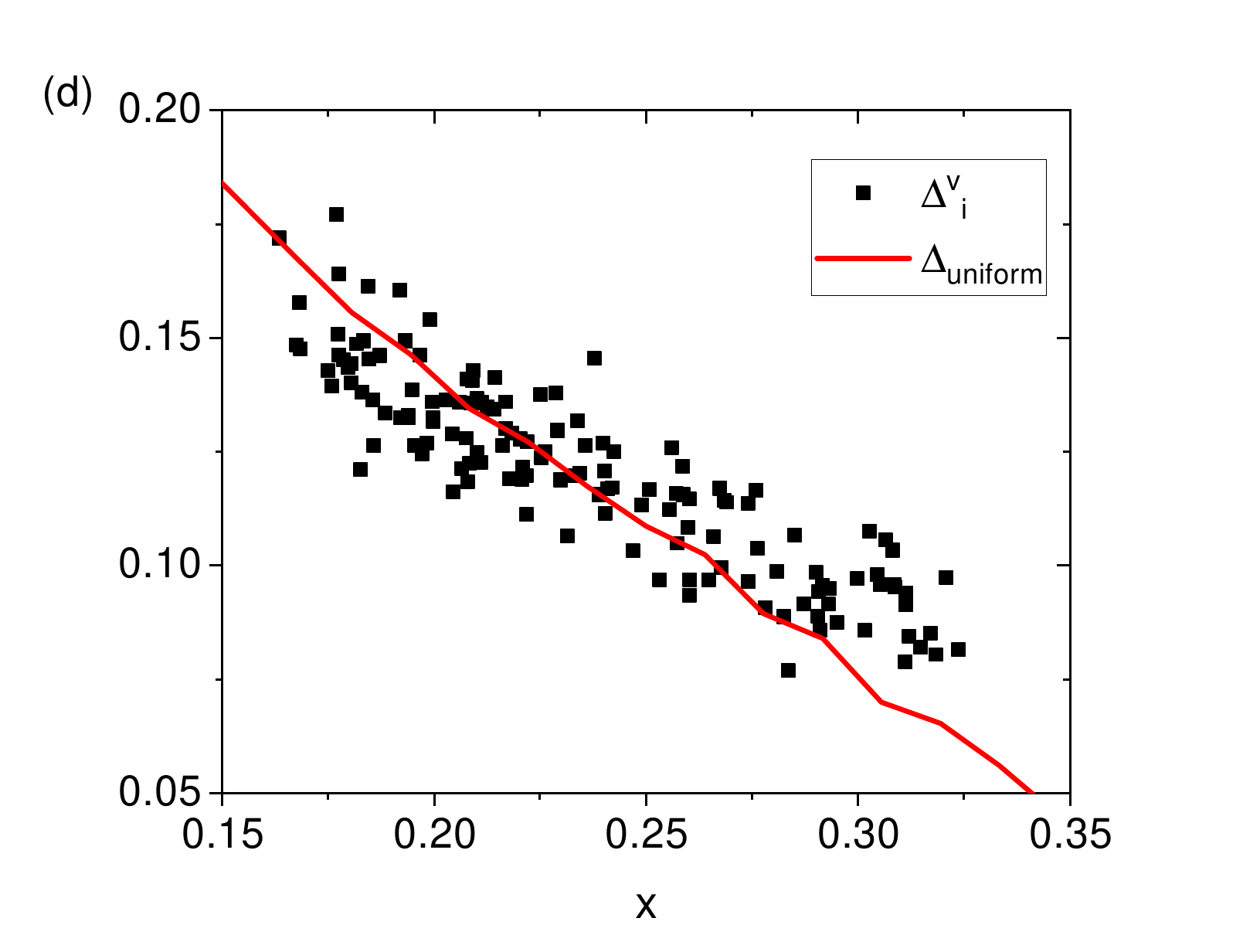}
\caption{The correlation between the optimized paring amplitude centered on site $i$, namely $\Delta^{v}_{i}$, and the local hole density. Here we set $V/t=1$ and the average hole concentration is $x=0.236$. The calculation is done on a $12\times12$ lattice with periodic-antiperiodic boundary condition. $\Delta^{v}_{i}$ is measured in unit of $t^{v}_{1,1+x}$, the first NN hopping parameter in the $x$-direction. The red line marks the doping dependence of the pairing amplitude in the clean system.}
\end{figure}  

We find that while the disorder potential can induce substantial inhomogeneity in the magnitude of the pairing amplitude $\Delta^{v}_{i,j}$, its spatial average is essentially unchanged. To be more quantitative, we plot the dependence of the spatially averaged pairing amplitude as a function $V/t$ for $x=0.236$ in Fig.6. It is found that the spatial average of $\Delta^{v}_{i,j}$ increases gently with the increase of the disorder strength $V/t$. This is an expected result from the above scenario by noting the following two facts. First, the average hole density is fixed when we tune the disorder strength. Second, the doping dependence of the pairing amplitude in the clean system is concave( see Fig.1). This result however, does not imply that the superconductivity of the disordered system would become more robust with the increase of the disorder strength. To characterize the superconductivity in the disordered system, we have calculated the off-diagonal-long-range-order(ODLRO) in the presence of the disorder potential. The ODLRO is defined as follows
\begin{equation}
F^{2}=\frac{1}{N}\sum_{i}\langle \hat{\Delta}_{i+\mathbf{R}}^{\dagger}\hat{\Delta}_{i}\rangle
\end{equation} 
in which $\mathbf{R}$ denotes the largest distance on the $12\times12$ lattice, $\hat{\Delta}_{i}$ is the pairing field centered on site $i$. It is defined as
\begin{equation}
\hat{\Delta}_{i}=\frac{1}{4}[\hat{\Delta}_{i,i+x}+\hat{\Delta}_{i,i-x}-\hat{\Delta}_{i,i+y}-\hat{\Delta}_{i,i-y}]
\end{equation}
in which
\begin{equation}
\hat{\Delta}_{i,j}=\hat{c}_{i,\uparrow}\hat{c}_{j,\downarrow}+\hat{c}_{j,\uparrow}\hat{c}_{i,\downarrow}
\end{equation}
is the paring field on the bond between site $i$ and $j$.

\begin{figure}
\includegraphics[width=8cm]{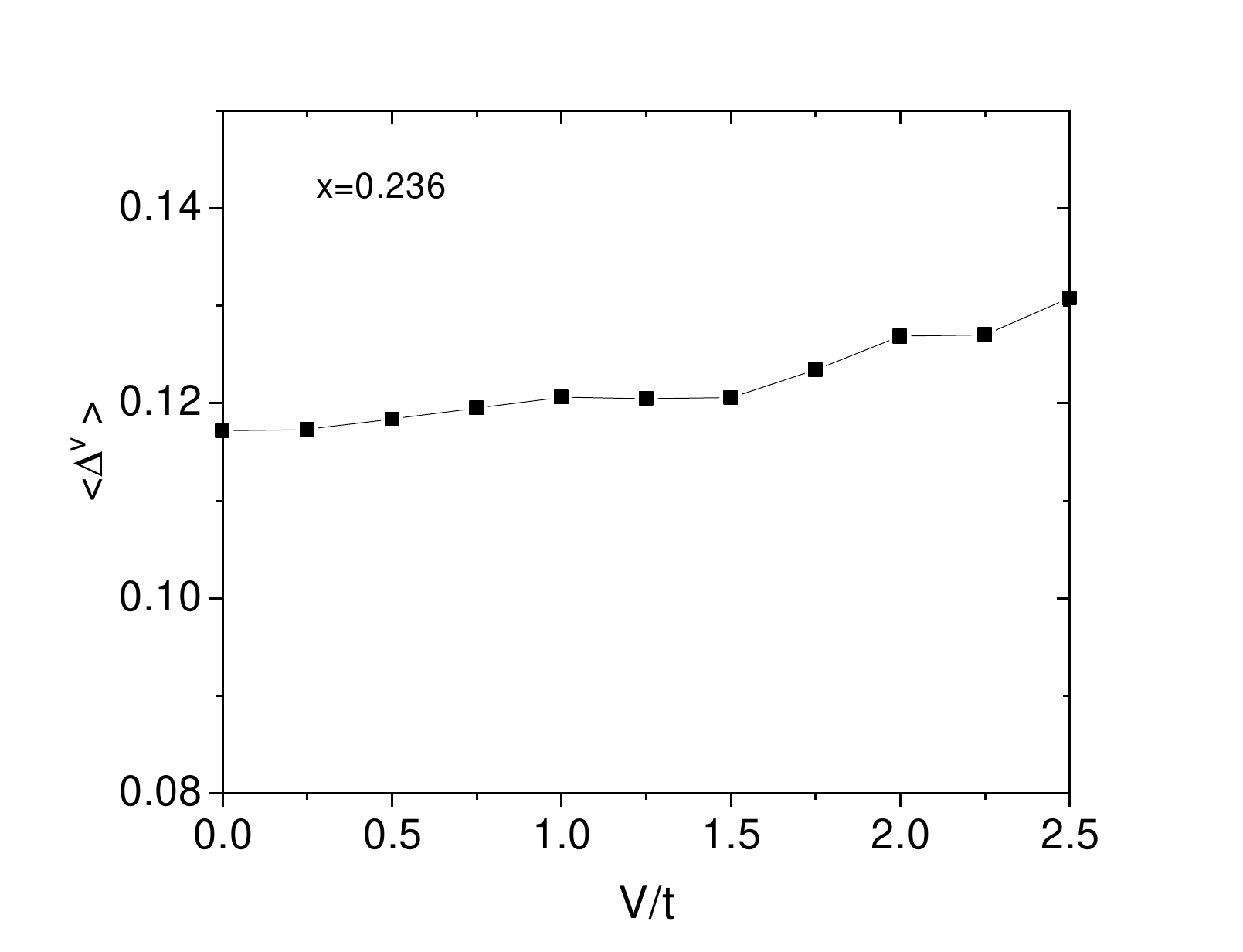}
\caption{The evolution of the spatial averaged pairing amplitude with the disorder strength $V/t$ at $x=0.236$. The calculation is done on a $12\times12$ lattice with periodic-antiperiodic boundary condition. The pairing amplitude is measured in unit of $t^{v}_{1,1+x}$, the first NN hopping parameter in the $x$-direction.}
\end{figure}  

The evolution of $F$ with the disorder strength is shown in Fig.7 for $x=0.236$. The ODLRO is found to decrease monotonically with $V/t$. However, the level of the reduction is rather small, amounting to only about 13 percent of its clean limit value at the strongest disorder potential of $V/t=2.5$. Such a disorder potential is already more than 8 times stronger than the bare Heisenberg exchange coupling $J$. In the plain BdG treatment, the d-wave superconductivity is totally suppressed by such a strong disorder. This result emphasizes again the additional robustness of the d-wave superconducting pairing aided by the strong correlation effect in the $t-J$ model. 

\begin{figure}
\includegraphics[width=8cm]{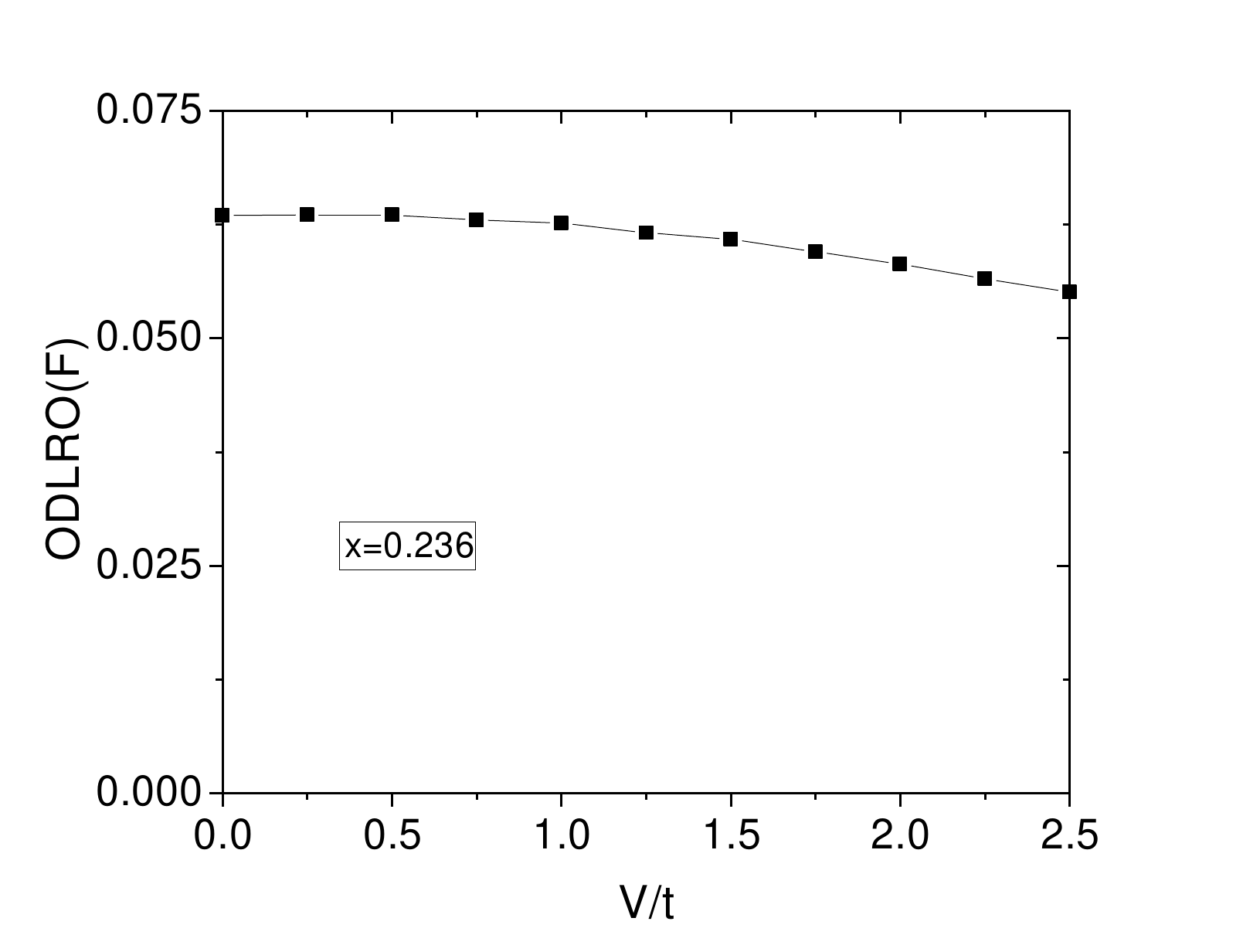}
\caption{The evolution of the ODLRO(F) with the strength of the disorder potential at $x=0.236$. The calculation is done on a $12\times12$ lattice with periodic-antiperiodic boundary condition.}
\end{figure}  

To have a complete picture of the disorder effect in the 2D $t-J$ model, we have mapped out the whole doping-disorder strength phase diagram, which is shown in Fig.8a for the spatial average of the pairing amplitude $\langle |\Delta^{v}|\rangle$ and Fig.8b for the ODLRO $F$. Similar to what is found for $x=0.236$, the average pairing amplitude is seen to increase gently with the increase of the disorder strength at all doping concentration. The ODLRO on the other hand is found to decrease gently with the increase of the disorder potential at all doping concentration. However, we find that the reduction in the ODLRO never exceed 20 percent of its clean limit value even at a disorder strength that is more than 8 times stronger than the Heisenberg exchange coupling $J$. This remarkable robustness of the d-wave pairing is thus a genuine characteristic of the 2D $t-J$ model and is surely beyond the BCS theory description. 

\begin{figure}
\includegraphics[width=8cm]{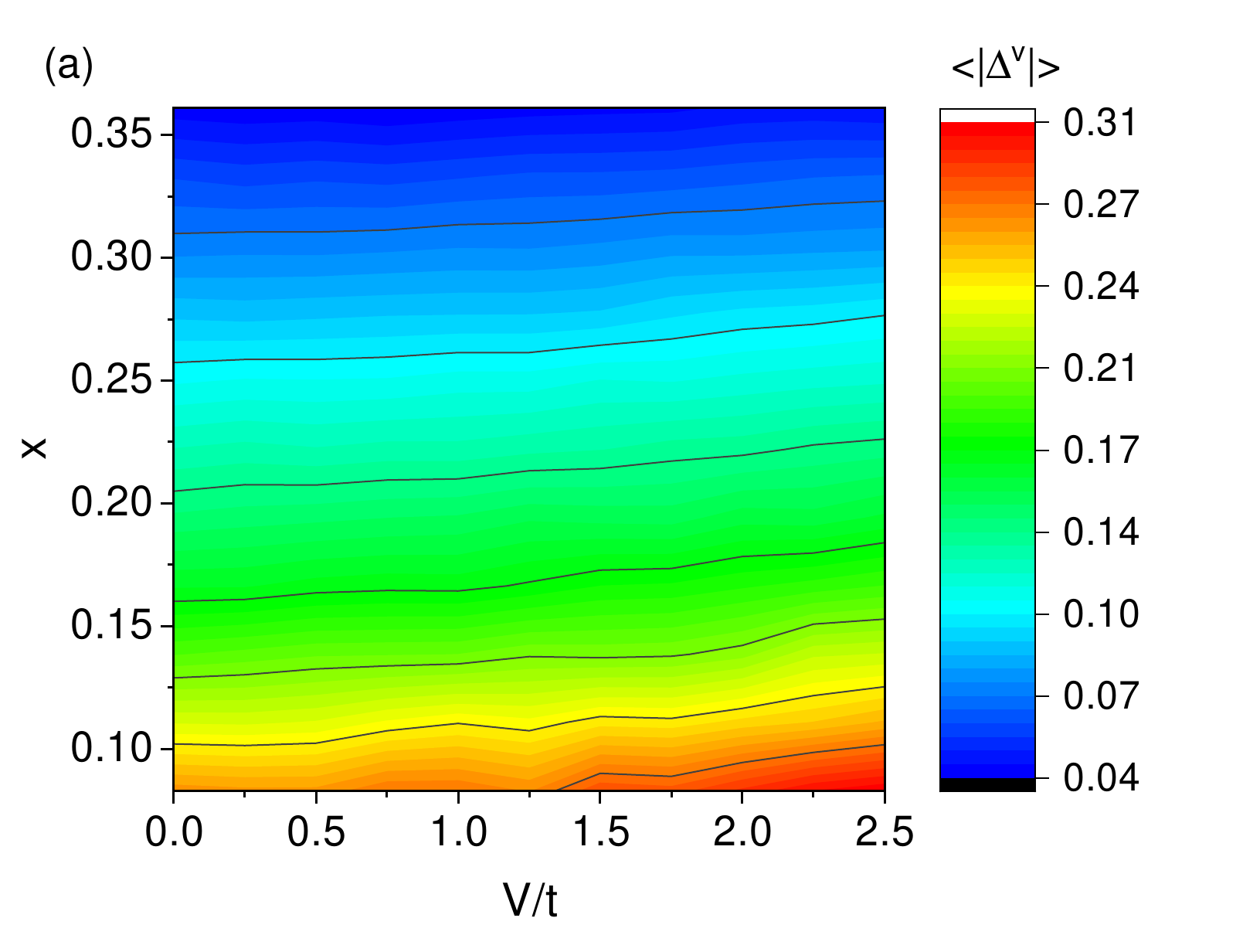}
\includegraphics[width=8cm]{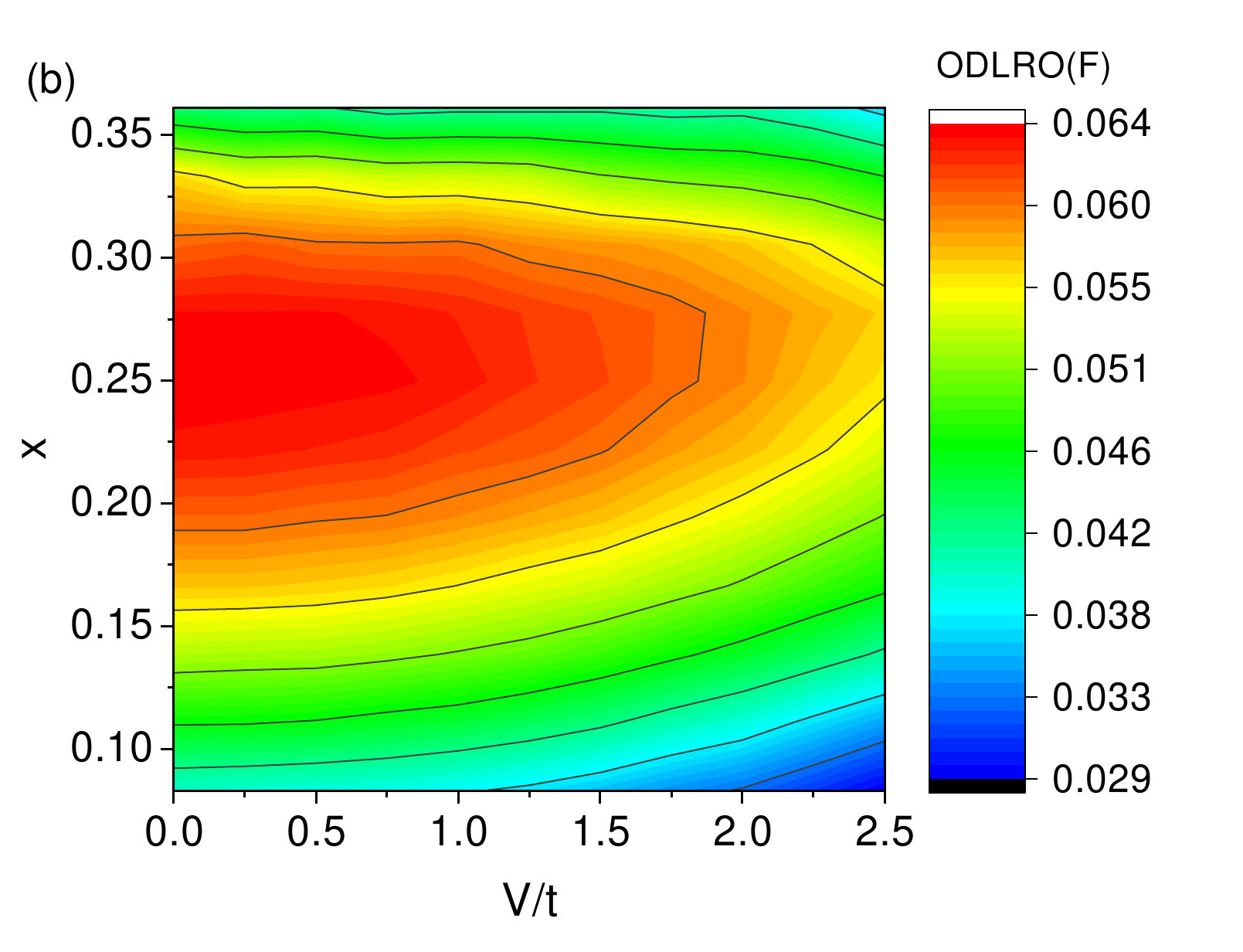}
\caption{The doping-disorder strength phase diagram of the 2D $t-J$ model obtained from our variational optimization. (a)the spatial average of the pairing amplitude $\langle |\Delta^{v}|\rangle$. (b)the ODLRO. The calculation is done on a $12\times12$ lattice with periodic-antiperiodic boundary condition. The pairing amplitude is measured in unit of $t^{v}_{1,1+x}$, the first NN hopping parameter in the $x$-direction.}
\end{figure}  

A quantity of more experimental relevance is the superfluid density $\rho_{s}(0)$ of the disordered system. However, unlike the off-diagonal-long-range-order (ODLRO) calculated above, the superfluid density is not a ground state property. As a dynamical property of the system, the calculation of $\rho_{s}(0)$ involves the information of the excitation behavior of the system and is currently beyond the reach of the variational Monte Carlo approach adopted here. Of course, if we relax the no double occupancy constraint and work at the level of the mean field approximation, the superfluid density can certainly be calculated. However, as we have emphasized in this work, the no double occupancy constraint is crucial to arrive at our conclusion. In recent years, variational Monte Carlo algorithm has also been developed to calculate the dynamical properties of the 2D $t-J$ model\cite{Li1}. However, in the absence of the translational symmetry, as is the case in this study, such a computation becomes too expensive to be conducted. 

An alternative but equivalent way to calculate the superfluid density of the system is to study the response of the ground state to the phase twist in the boundary condition induced by the insertion of a gauge flux\cite{Scalapino}. More specifically, in the natural unit in which $e=c=\hbar=1$ we have
\begin{equation}
\frac{\rho_{s}(0)}{\pi}=\lim_{L_{y}\rightarrow \infty}\lim_{L_{x}\rightarrow \infty}\frac{1}{L_{x}L_{y}}\frac{\partial^{2}E}{\partial A_{x}^{2}}
\end{equation} 
Here $L_{x}$ and $L_{y}$ are the linear size of the system in the $x$ and $y$-direction. $A_{x}$ is a uniform vector potential in the $x$-direction coupling to the electron through the Peierls substitution in the hopping integral of the $t-J$ model
\begin{equation}
t_{i,j}\rightarrow t_{i,j}e^{iA_{x}(i_{x}-j_{x})}
\end{equation}
in which $t_{i,j}$ denotes the hopping integral between site $i$ and $j$. $i_{x}$ and $j_{x}$ are the $x$-component of their lattice coordinates. In principle, Eq.37 can be used to estimate the superfluid density through a variational optimization of the ground state energy of the system in the presence of the vector potential $A_{x}$. However, we note that on a finite system a stable variational optimization is possible only when the considered variational ansatz satisfy the closed-shell condition. On the other hand, such a condition is not always guaranteed when the vector potential $A_{x}$ is varied continuously. In fact, the paramagnetic response of system is just carried by the cloud of quasiparticle excitation whose energy would drift linearly with $A_{x}$ and may cross the fermi level when we increase $A_{x}$. A possible way to solve such a technical problem is under consideration.

\section{Conclusions and discussions}
In this work, we have studied the fate of the d-wave pairing in the 2D $t-J$ model in the presence of the disorder potential from a variational perspective. The results can be summarized as follows. We find that the d-wave pairing in the 2D $t-J$ model is remarkably more robust against the disorder effect than that in a conventional d-wave BCS superconductor. More specifically, we find that the phase structure of the d-wave pairing is well preserved both locally and globally even at the strongest disorder strength that we have simulated, which is more than 8 times stronger than the Heisenberg exchange coupling in the $t-J$ model. This is very different from the situation in a disordered d-wave BCS superconductor, in which the destructive interference of the pairing amplitude caused by impurity scattering may even lead to frustration in the phase of the pairing order parameter at large scale\cite{Lizx,Berg}. The spatial average of the pairing amplitude is found to increase gently with the the increase of the disorder strength. At the same time, we find that the reduction in the ODLRO never exceed 20 percent of its clean limit value. We find that these conclusions hold at all doping level across the superconducting dome and is a robust property of the 2D $t-J$ model.

We find that the disorder potential has its most significant effect on the holon condensate amplitude $\bar{b}_{i}$. On the other hand, the spatial variation in the spinon chemical potential $\mu^{v}_{i}$ is found to be an order of magnitude weaker than that in the disorder potential $\mu_{i}$. This implies that the spinon system is essentially immune to the disorder potential. We find that the spatial modulation in the d-wave pairing amplitude can be understood as a secondary effect resulted from the modulation in the local hole density induced by the disorder potential. This is very different from the situation in the BCS scenario, in which the disorder potential acts directly on the electron participating in the d-wave Cooper pairing. The drastic suppression of the disorder effect on the d-wave paring of the $t-J$ model can thus be attributed to the spin-charge separation mechanism, through which the d-wave RVB pairing of the charge neutral spinon gains its robustness against the action of the disorder potential.

The remarkable robustness of the d-wave pairing in the $t-J$ model is clearly at odds with the observations in the overdoped cuprate superconductors, in which strong evidences for the fragility of d-wave pairing against the disorder effect have been found recently\cite{Lizx,Kim}. The contrasting behavior of the underdoped and the overdoped cuprates, namely, the remarkable robustness of the d-wave pairing in the former and its fragility in the latter, implies that the d-wave pairing on both sides of the phase diagram are of different nature. Here we propose that unlike the situation in the underdoped cuprates, in which the d-wave pairing should be understood as the RVB pairing between charge neutral spinons in a doped Mott insulator background, the d-wave pairing in the overdoped cuprates should be better understood as the more conventional BCS pairing between electrons in a fermi liquid metal background. This proposition offers a natural explanation for the observed non-monotonic doping dependence in $\rho_{s}(0)$. More specifically, since the RVB pairing in the underdoped cuprates is robust against the disorder effect, $\rho_{s}(0)$ should be dominated by the density of mobile charge carriers and should thus increase monotonically with $x$, as what we expect for a doped Mott insulator. As the electron becomes increasingly more itinerant in the overdoped cuprates, a description in terms of the conventional fermi liquid become increasingly more relevant. As a result, the d-wave BCS pairing between electrons becomes increasingly more fragile against impurity scattering. $\rho_{s}(0)$ is thus expected to decrease with increasing doping above some critical doping $x^{*}$ as a result of such transmutation in the nature of pairing.

One inference that can be drawn from the above proposition is that the pseudogap end point $x^{*}$ where $\rho_{s}(0)$ reaches its maximum should be understood as the transition point between a doped Mott insulator and a more conventional fermi liquid metal. Indeed, evidences in support of the abrupt enhancement of electron itinerancy at $x^{*}$ have been reported in many recent measurements\cite{Tallon,Davis,Minola,Chen}. However, we note that a Mott transition at a general incommensurate filling is still not a well accepted notion, even though such a transition has been claimed for the Hubbard model in a previous DMFT study\cite{Sordi}. In real cuprate superconductors, a finite doping Mott transition may well be driven by other mechanisms. One such possibility is the positive feedback between the enhancement of electron itinerancy and the screening of the Coulomb repulsion. More specifically, the enhancement of electron itineracy will strengthen the screening of the Coulomb repulsion between the electron, which will again enhance the electron itineracy further. Such a positive feedback loop may be responsible for the abrupt breakdown of a doped Mott insulator. This may be accompanied by the simultaneous reduction of the charge transfer gap with the increase of the hole concentration. In real cuprate superconductors, the doped hole mainly occupy the oxygen $2p$ orbital. The increase of the hole density will reduce the charge transfer gap between the oxygen band and the upper Hubbard band on the copper site if the repulsive potential between neighboring copper site and oxygen site is non-negligible.  

We note that no matter what is ultimate origin of such a potential finite doping Mott transition, it is surely out of the Landau paradigm of quantum phase transition involving spontaneous symmetry breaking order. The key quantity that governors such a transition is the electron itinerancy or electron coherence, which seems to decrease continuously when we approach $x^{*}$ from above. It is interesting to note that $x^{*}$ is also the place where the pesudogap behavior emerges. This implies that the Mottness of electron and the RVB pairing between the charge neutral spinons in such a doped Mott insulator is at the heart of the mystery of the enigmatic pseudogap phenomena. It is also interesting to see how such evolution of electron itineracy would lead to the observed strange metal behavior\cite{Hussey}.

\begin{acknowledgments}
We acknowledge the support from the National Natural Science Foundation of China(Grant No.12274457).
\end{acknowledgments}

\end{document}